# WSC-Trans: A 3D network model for automatic multi-structural segmentation of temporal bone CT

Xin Hua, *Student Member IEEE*, Zhijiang Du, Hongjian Yu, Jixin Ma, Fanjun Zheng, Cheng Zhang Qiaohui Lu, Hui Zhao

***Abstract*— Cochlear implantation is currently the most effective treatment for patients with severe deafness, but mastering cochlear implantation is extremely challenging because the temporal bone has extremely complex and small three-dimensional anatomical structures, and it is important to avoid damaging the corresponding structures when performing surgery. The spatial location of the relevant anatomical tissues within the target area needs to be determined using CT prior to the procedure. Considering that the target structures are too small and complex, the time required for manual segmentation is too long, and it is extremely challenging to segment the temporal bone and its nearby anatomical structures quickly and accurately. To overcome this difficulty, we propose a deep learning-based algorithm, a 3D network model for automatic segmentation of multi-structural targets in temporal bone CT that can automatically segment the cochlea, facial nerve, auditory tubercle, vestibule and semicircular canal. The algorithm combines CNN and Transformer for feature extraction and takes advantage of spatial attention and channel attention mechanisms to further improve the segmentation effect, the experimental results comparing with the results of various existing segmentation algorithms show that the dice similarity scores, Jaccard coefficients of all targets anatomical structures are significantly higher while HD95 and ASSD scores are lower, effectively proving that our method outperforms other advanced methods.***

## I. INTRODUCTION

COCHLEAR implantation is currently the most effective treatment for profound deafness [1-3]. Prior to surgery, clinicians need to determine the surgical location and the path of electrode implantation using the patient's preoperative CT. Because the cochlear nerve is small compared to other biological tissues, it is difficult to identify in CT, takes a long time to identify, and requires a high level of skill and experience. The location of the cochlea and other biological tissues in the temporal bone and the relationship between the cochlea and the facial nerve and other tissues in space are shown in Fig 1. Therefore, it is a major challenge to segment the cochlea and nearby biological tissues in the temporal bone quickly and accurately.

Early approaches to medical image segmentation usually relied on techniques such as template matching techniques, edge detection, statistical shape models, and traditional machine learning [4]. These methods have achieved good results to some extent, however, because of the issues with blur, noise, and low contrast in medical images [5], image feature representation is difficult and requires manual annotation resulting in inefficiency, and medical image segmentation is still one of the most challenging topics in the field of computer vision. As deep learning techniques improves by leaps and bounds, convolutional neural networks (CNNs) have successfully achieved multi-level feature representation of images, thus becoming the hottest research topic in the field of image processing and computer vision. The insensitivity of CNN for feature learning to image noise, blur, contrast, etc. makes it outstanding in medical image segmentation.

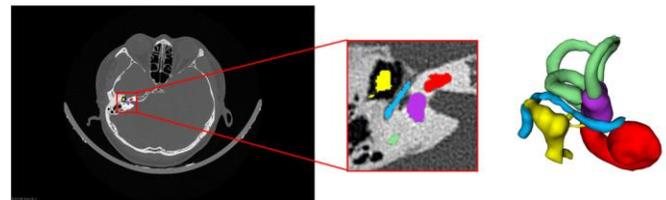

Fig1. In the temporal bone CT shows cochlea(red), facial nerve(blue), ossicles(yellow), vestibule(purple), and semicircular canal(green) and show the spatial relationship between organizations in three dimensions

U-net [6] is the most popular CNN-based semantic segmentation method in the field of medical image segmentation. Since the Unet network has good performance and simple structure at the same time, many researchers have proposed Unet3+ [7], nnUnet [8], KiUnet [9], R2UNet [10], QAP-Net [11], Double-Unet [12] and other models using the core idea of Unet, which have good performance in semantic

This work involved human subjects in the study. All ethics and protocols were approved by the Medical Ethics Committee of the Sixth Medical Center of the General Hospital of the People's Liberation Army of Chinese.( *Corresponding author:Hongjian Yu*) This project was supported by the National Key Research and Development Program of China under Grant No. 2019YFB1311800, "Research on Minimally Invasive Cochlear Implant Robot System".

Xin Hua, Jixin Ma studied in China, Harbin Institute of Technology, School of Mechanical and Electrical Engineering(e-mail: huaxin@stu.hit.edu.cn, 1180801226@stu.hit.edu.cn), Zhijiang Du, Hongjian Yu worked in China, Harbin Institute of Technology, School of Mechanical and Electrical Engineering(e-mail: duzj01@hit.edu.cn, yuhongjian@hit.edu.cn) Chen Zhan, Zheng Fanjun studied in China, the People's Liberation Army General Hospital(email: zhch2337@163.com, zhengfanjun1989@126.com), Qiaohui Lu worked in China, the People's Liberation Army General Hospital, the sixth medical center of emission diagnosis(e-mail: luqiaohui465@126.com), Hui Zhao worked in China, the People's Liberation Army General Hospital, the Department of Otolaryngology, Head and Neck Surgery Medical(e-mail: huizhao@yeah.net).

segmentation tasks. Since medical images (CT, MRI, etc.) are all 3D, some organs, or tissues in image segmentation where contextual information has a greater impact on segmentation accuracy can achieve better results using 3D segmentation. Researchers have proposed 3D U-Net [13], V-Net [14], 3D AGSE-VNet [15] and other models for 3D segmentation of medical images. Although CNN has excellent performance in local feature extraction, the ability of CNN to capture the global feature representation of the image is still lacking, and each convolutional kernel in CNN is limited by the size of its own kernel and the pace size can only focus on its own image features, lacking a large range of image feature fusion, which in turn affects the overall image segmentation effect One of the most intuitive solutions is to expand the perceptual field, but this would disrupt the operation of the pooling layer [16]. To overcome these drawbacks, researchers have proposed several solutions including the use of null convolution [17], deformation convolution [18], and Dynamic Convolution [19] to expand the feature focus. Although it is a certain improvement of model capability but there are some limitations.

Transformer has been successful in recent years in the fields of 2D image classification and natural language processing. Transformer is designed to capture relationships between arbitrary positions in a sequence as well as model long-range correlations in sequence-to-sequence tasks. This architecture is suggested purely based on self-attention [20] and does not require convolution at all in extracting features. As a popular model dominating natural language processing tasks, transformer is now being introduced to computer vision tasks with promising results in target detection, image classification, and semantic segmentation tasks. Vision Transformer (ViT) [21] segments images into blocks and uses transformer to correlate between these blocks Modeled as sequences. By segmenting each image into Patches with positional embeddings, ViT creates a series of tokens and then extracts parameterized vectors as visual representations using Transformer Blocks. Because of the Self-Attention mechanism and the Multi-Layer Perceptron (MLP) structure, Vision Transformer can reflect intricate spatial transformations and long-range feature dependencies to obtain a global feature representation. However, Vision Transformer can ignore local feature details. Therefore, some works suggest a tokenization module or use CNN feature maps as input token in order to gather feature proximity information. the use of window attention mechanism in Swim Transormer [22] improves the computational efficiency more and shifts the window also makes the adjacent windows interact with each other and the upper and lower layers The connection is established, thus achieving a global modeling capability in disguise. However, these methods still do not fundamentally solve the relationship between local modeling and global modeling.

1.Design and development of a semantic segmentation network model for accurate segmentation of the temporal bone and nearby 3D anatomical structures with extremely complex minutiae.

2. In the encoding part and the decoding part, we use Dense method and Res method to design a new feature extraction and response module for feature extraction and response respectively.

3. Feature extraction was carried out in the Dense module in the coding part, and feature recovery was carried out in the Res module in the decoding module. In the advanced semantic part, window attention，shift window attention and spatial attention mechanism were proposed to obtain more relevant feature information and corresponding location information.

4. The CT images of temporal bone were used to evaluate the experimental results. The evaluation results indicate that the proposed model outperforms the previous advanced methods, thereby demonstrating the efficacy of our approach.

## II. RELATED WORKS

### A. U-net and its variants

Res-Unet [23] and Dense-Unet [24] are inspired by residual linking and dense linking, respectively, and use each submodule of Unet with residual linking and dense linking, respectively. U-net++ [25] is equivalent to stitching together four U-net networks of different depths through multiple skip connections. Short connections are used in the network to realize the feedback operation of the whole network, allowing training to proceed, and long connections are used to connect the input image information before and after, reducing the information loss caused by downsampling during encoding. resUnet++ [26] can be seen as a model designed by combining U-net++ and the ResUnet model, and MultiResUNet [27] designs a MutiRes module that stitches together three convolutional results as a combined feature map and introduces residual connections, where the feature map passes from the encoder through the chain of convolutional layers with residual connections during jump connection and then connects to the decoder features. DC-Unet [28] is designed to address the problem of invalid spatial features in MultiResUnet [27] by using 2 paths of Three 3x3 convolutional layers are used instead of the residual connection in MultiRes module, and finally the results of these two pathways are linearly summed to output. the design in U2-Net [29] uses RSU modules instead of encoding modules as well as decoding modules, and each RSU module itself is a small-sized Unet, and a large Unet-like network model is constructed with multiple small Unets. To increase the multi-scale capability, DobuleU-Net [30] is equivalent to a combination of two UNet network structures connected and uses ASPP [31] spatial pyramidal pooling to capture contextual information.M-Net [32] adds input features of different scales at different levels to capture multi-level visual details through a series of up-sampling and down-sampling layers. And the output features of different levels are fused to obtain segmentation results. PWD-3DNet [33] adopts a similar DenseUnet structure in the encoding part, and directly uses B-spline for up-sampling in the decoding part and fuses the output features of the decoding part of different levels after up-sampling them to a uniform size to obtain segmentation results.

### B. Transformer and its variants

The Swim-Unet [37] model uses a layered Swin Transformer with offset window as the encoder. And a symmetric Swin Transformer-based decoder with a patch extension layer is designed to perform the up-sampling operation to recover the spatial resolution of the feature map. The tokenized image blocks from the encoder are fed into the decoder in the same layer using jump connections for local global semantic feature learning. In the Cotr [35] model, researchers designed the deformable Transformer (DeTrans) with a deformable self-attentive c to reduce the computational and spatial modeling of long-distance dependencies of multi-scale and high-resolution feature maps complexity. Swin Transformer is applied to both the encoder and decoder in DS-TransUNet [36] to extract coarse-grained and fine-grained feature representations at different language scales using a dual-scale encoder sub-network based on Swin Transformer. A module named Transformer Interactive Fusion (TIF) is designed, which uses a self-attentive mechanism to efficiently establish global dependencies between features at different scales. Transfuse [37] uses Transformer for feature extraction and the decoder part uses the progressive up sampling (PUP) structure used in SERT [38]. The model uses channel attention for the image features of the encoder and spatial attention for the CNN Branch. Then various operations such as convolution, multiplication, splicing, and residuals are performed to achieve feature fusion of the two branches. Finally, the segmentation results are output by up-sampling and attention-gated skip-connection. TransBTS [39] To obtain local 3D contextual information, the encoder first extracts 3D spatial feature maps using 3D CNNs. At the same time, the feature mapping is carefully modified to input the tokens into the Transformer for global feature modeling. The decoder performs progressive up sampling using the features embedded in the Transformer to predict the segmentation map in detail. mCTrans [40] embeds multi-scale convolutional features into a sequence of tokens and performs intra-scale and inter-scale self-attention. Also proposed are learnable agent embeddings, which utilize self-attention and cross-attention to model semantic relations and feature augmentation, respectively. nnFormer [41] uses Volume-based Multi-head Self-attention (LV-MSA) and Global Volume-based Multi head Self-attention (GV-MSA) to jointly build multi-level feature pyramids for local as well as global feature representation, which can effectively reduce computational effort while improving accuracy. To expend the existing architecture, medT [42] presented a Gated Axial-Attention model by adding more control mechanisms to the self-attention module. Transclaw u-net [43] uses a model structure similar to Claw U-Net [44] combined with the Transformer mechanism and uses the Transformer mechanism for feature extraction at the bottom layer. The extracted features are unsampled and added to the decoding part respectively, and finally the same layer encoding features, up sampling features, and decoding features are combined to obtain the final segmentation results.

### III. MATH

The main structure of the model we designed is shown in Figure 2. This work is mainly inspired by TransBTS[39] model, but the model structure is different .TransBTS establishes a number of similar ResNet structures in the encoding and decoding part to ensure image feature acquisition, and uses VIT structures in the bottom part of the network for feature learning. The model we designed uses the WSC module at the bottom, which consists of window attention, shift window attention, and channel attention in series.

In the coding part, the Dense module is used to extract image features, which is mainly used to transform the input images into features that can be processed by the network. Convolutional networks can better retain accurate location information of features, and convolutional operations can provide high-resolution underlying features [41]. However, considering the characteristics of medical images, the DensBlocke module is used to extract as many low-level features as possible to preserve the images. and Skip connect is used to transfer the information features to the decoding part, so that the shallow information and contour information in medical images can be effectively combined with the features in the decoding part to improve the segmentation effect.

$$\begin{aligned} x_1 &= Conv(ReLu(GroupNormal(x))) \\ x_2 &= Conv(\mathrm{Re}Lu(GroupNormal(x_1))) \quad (1) \\ y &= Conv(Concat(x_1, x_2)) \end{aligned}$$

The Pooling layer is usually used for down-sampling. The field of view is limited by kernel size and stride, and the maximum feature value can only be selected in a limited range. In the down-sampling process, although the convolution layer retains the location information of features, only the maximum value of features can be known after Pooling layer. But the location information is not effectively retained, and sometimes some features appear multiple times. Since Pooling only retains the maximum value, even if a feature appears multiple times, only a small amount of relevant information is retained, that is, the intensity information of the same feature is lost. Therefore, used in the process of encoding module in part down sampling in ViT Patch Merging method, using this method in the down sampling at the same time, by improving the channels are reserved to the number of ways to feature information position relationship between retention characteristics at the same time, in order to avoid the traditional image pixel information loss in the process of sampling, to the best degree of retain the original information. As the input characteristics of down-sampling for the 3D model were C×H×W×D to 8C×H/2×W/2×D/2, the number of channels changed dramatically, so Conv1×1 was used to change the channel to 2C.

An up-sampling convolution based on learning can effectively restore feature information. In the decoding part, the similar Res module can retain image features to the maximum extent during image recovery. Skip-connect is used to transfer the features extracted from the coding part of the same layer to the decoding part. In the decoding part of each layer, the features extracted from the decoding part of the same layer are utilized as well as the high-order semantic information. The corresponding equation is shown:

$$\begin{aligned} x_1 &= Conv(ReLu(GroupNormal(x))) \\ x_2 &= Conv(\mathrm{Re}Lu(GroupNormal(x_1))) \quad (2) \\ y &= Conv(x_1 + x_2) \end{aligned}$$

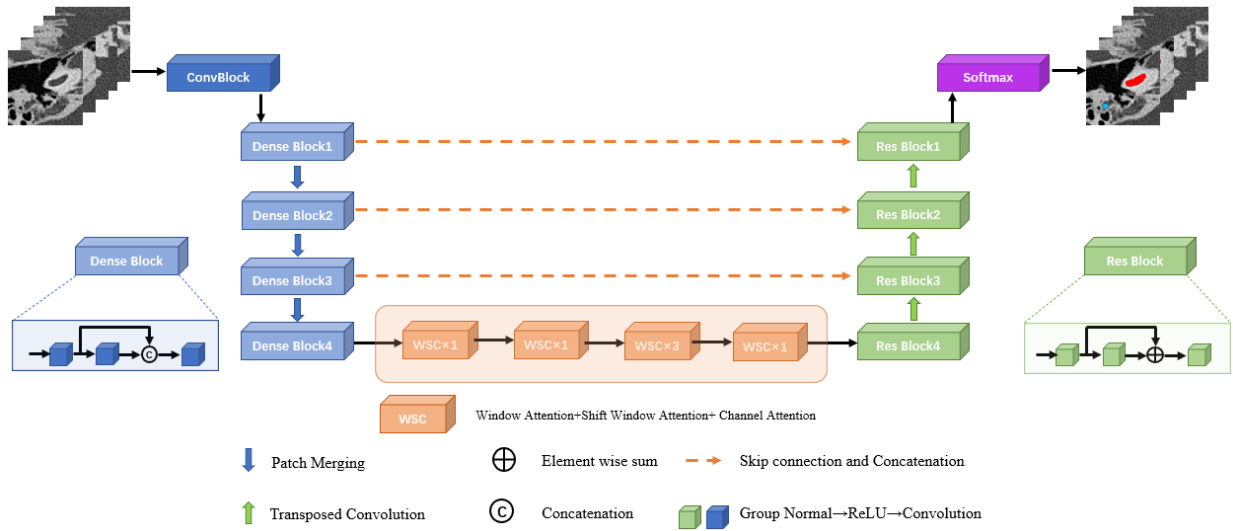

Fig4 Architecture of the proposed WSC-Trans segmentation algorithm, In the In the encoding part of the image, first use ConvBlock to boost the number of channels, then use DenseBlock to extract the shallow image features and pass the features to the decoding part of the same layer, and Patch Merging for down sampling, and use WSC module for advanced semantic feature extraction at the bottom layer, and use Transposed Convolution for up sampling in the decoding part, and recover the image features by ResBlock with the features passed to the encoding part of the same layer.

Inspired by Davit model [45], the spatial attention mechanism is combined with channel attention mechanism to obtain comprehensive information of features, and window attention and shift window attention in Swim Transformer are used as the spatial feature acquisition module. Window attention can establish a linear relationship with patches to reduce the calculation, but it is a local attention mechanism that enables feature extraction in a wide range by connecting each patch with hyperbolic window. Use a combination of window attention and hyperbolic window attention to expand the local information search into global information extraction. Channel attention is used to capture important channel information and capture the required global information. We choose the channel attention model structure in Davit as shown in Figure 3, and not use the Token mechanism to further reduce the computational effort.

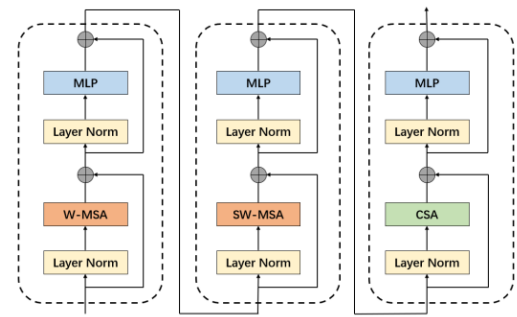

Fig4. WSC model structure, widow attention first, then shift window attention and finally channel attention, window attention and shift window attention to complete the spatial feature acquisition, channel window attention and shift window attention complete the spatial feature acquisition, and channel attention completes the channel feature acquisition.

WSC-Trans network related parameters are shown in Table I.

TABLE I
PARAMETERS OF WSC-Trans ALGORITHM

| Layer | Input size | Output size |
|---|---|---|
| ConvBlock | 1×96×96×96 | 32×96×96×96 |
| Dense Block1 | 32×96×96×96 | 32×96×96×96 |
| Down1 | 32×96×96×96 | 64×48×48×48 |
| Dense Block2 | 64×48×48×48 | 64×48×48×48 |
| Down2 | 64×48×48×48 | 128×24×24×24 |
| Dense Block3 | 128×24×24×24 | 128×24×24×24 |
| Down3 | 128×24×24×24 | 256×12×12×12 |
| Dense Block4 | 256×12×12×12 | 256×12×12×12 |
| WSC | 256×12×12×12 | 256×12×12×12 |
| Res Block4 | 256×12×12×12 | 256×12×12×12 |
| Up3 | 256×12×12×12 | 128×24×24×24 |
| Res Block3 | 128×24×24×24 | 128×24×24×24 |
| Up2 | 128×24×24×24 | 64×48×48×48 |
| Res Block2 | 64×48×48×48 | 64×48×48×48 |
| Up1 | 64×48×48×48 | 32×96×96×96 |
| Res Block1 | 32×96×96×96 | 32×96×96×96 |
| Softmax | 32×96×96×96 | 6×96×96×96 |

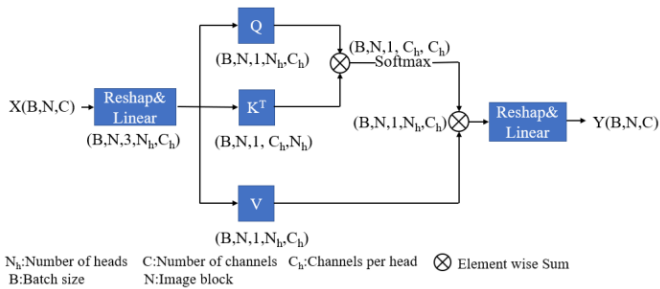

Fig3. Channel attention structure (CSA), the structure and vit in attention calculation in the same way only the difference lies in the calculation of the channel.

By combining spatial attention mechanism with channel attention mechanism, the accurate information of feature can be obtained better. The detailed structure of the model is shown in Figure 4.

## IV. EXPERIMENTS

### A. Materials

CT imaging data of temporal bone in 80 adult healthy individuals were provided by General Hospital of the Chinese

People's Liberation Army with resolution 1024x1024 and layer thickness from 207 to 306. CT voxels were all 0.23mm*0.23mm*0.33mm, and all temporal bone CTs were labeled by an experienced imaging physician to label the cochlea, vestibule, semicircular canal, facial nerve, and auditory tuberosity. The total number of pixels of the target in each CT is less than 0.01% of the overall number of pixels. To save computation the CTs of temporal bone parts are cropped to remove unnecessary influential parts, and the CT pixels are 96x96x96 after cropping, and due to the symmetry of human body, the number of materials is extended to 160, of which we randomly select 128 as the training set and 32 as the test set. The CT data of the training and test sets were randomly shifted along the axes [8.3%-10.4%], expanding the training data to 288 and the test data to 96, respectively..

### B. Loss function

Since we clearly know that the number of pixels to be identified for segmentation occupies less than 2% of the total number of pixels, and the ratio of the number of pixels for each segmentation target is extremely unbalanced, Dice Loss is chosen as the loss function due to its good performance in unbalanced samples.

### C. Implementation details

All models in this paper are implemented by pytorch, using Windows system, Intel(R) Xeon(R) Gold6240R CPU, TeslaA100 40G GPU, Batch Size of 4, 40 epochs of learning, learning rate of 0.001, AdamW as the accelerator where the relevant parameters beats = 0.9-0.999, eps=1e-8, and weight decay=0.01.

## V. RESULTS

### A. Experimental evaluations

In this experiment, four evaluation methods were chosen to evaluate the model segmentation effect, namely Dice similarity score (DSS), Jaccard similarity score (JSS), 95% Hausdorff Distance (HD), and Average Symmetric Surface Distance (ASSD), which are defined in relation to each other as follows. Where P represents the segmentation result predicted by the model, G represents the real label.

$$DSS(P,G) = \frac{2|P \cap G|}{|P| + |G|} \quad (3)$$

Jaccard similarity coefficient (Jaccard similarity coefficient) is mainly used to compare the similarity and difference between limited sample sets.

$$JSS(P,G) = \frac{|P \cap G|}{|P| + |G| - |P \cap G|} \quad (4)$$

HD95 due to the description of the maximum degree of mismatch between two-point sets, considering the exclusion of some outlier points to produce unreasonable data, the impact on data stability, in the selection of distance can cover 95% of the distance in the number of bits to ensure the stability of the data.

$$HD(P,G) = max(h(P,G), h(G,P))$$
$$h(P,G) = \max_{p \in P} \left\{ \min_{g \in G} \| p - g \| \right\} \quad (5)$$
$$h(G,P) = \max_{g \in G} \left\{ \min_{p \in P} \| g - p \| \right\}$$

ASSD mainly indicates the size of the distance between two image voxels.

$$ASD(P,G) = \sum_{p \in P} \min_{g \in G} d(p,g)/|P|$$
$$ASSD(P,G) = \{ASD(P,G) + ASD(G,P)\}/2 \quad (6)$$

### B. Segmentation results

We used 4-fold cross-validation, each model selected the model parameters with the best results in cross-validation for the validation set to obtain the results. The 2D and 3D segmentation effect of each model is shown in Figure 5 and Figure 6, where 3D Slicer is used to show the 3D segmentation effect.

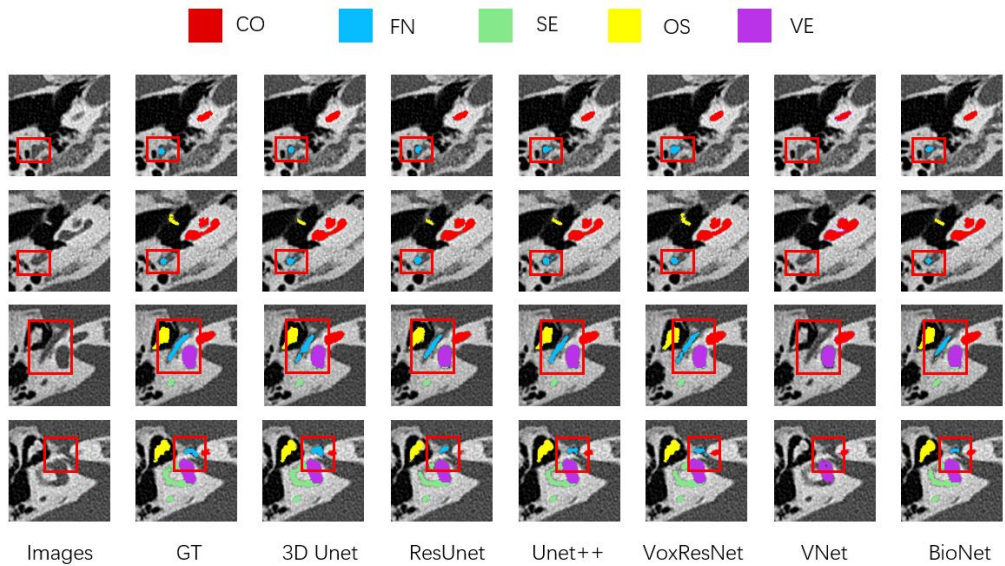

Images | GT | 3D Unet | ResUnet | Unet++ | VoxResNet | VNet | BioNet

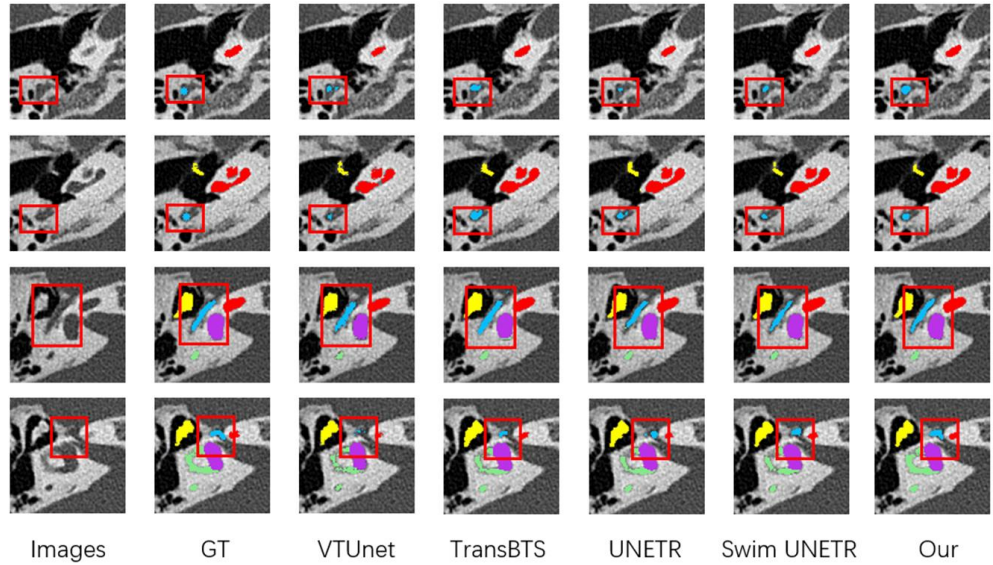

Fig5. Segmentation effect of cochlea(red), facial nerve(blue), ossicles(yellow), vestibule(purple), and semicircular canal(green) in different layers of CT of temporal bone segmentation results compared to several other advanced medical image segmentation models listed in this paper.

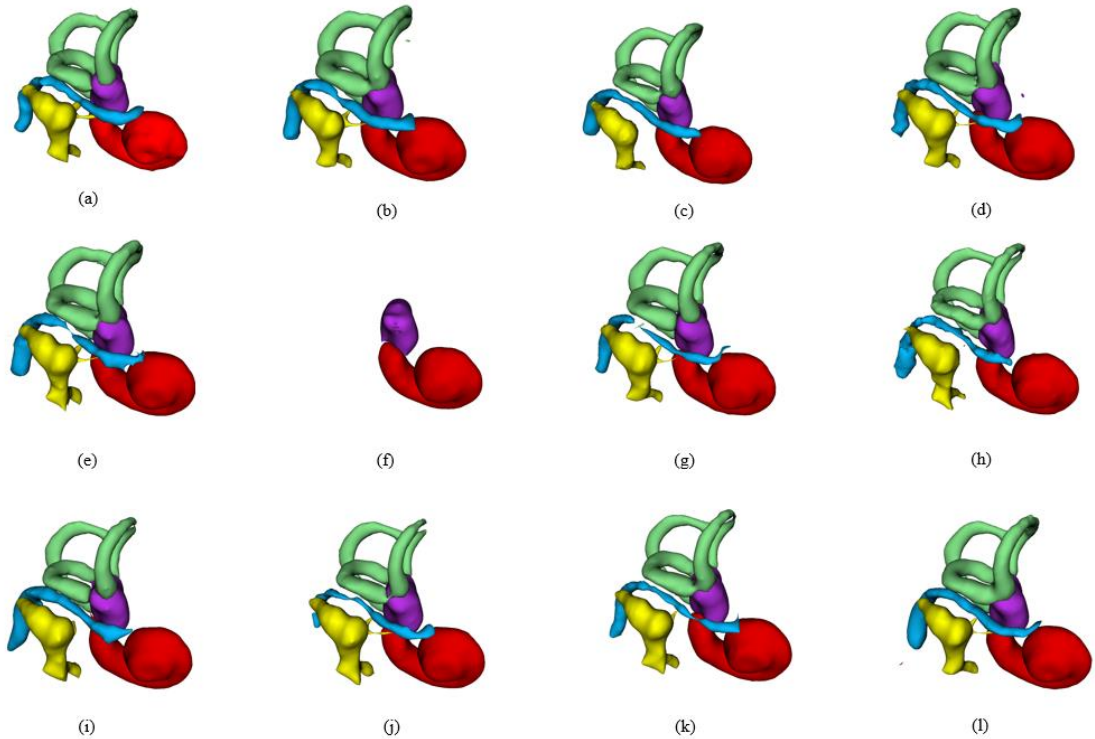

Fig6. (a)~(l) represent GT, 3D Unet, 3D ResUnet, Unet++, VoxResUnet, Vnet, 3D BioUnet,,VT-Unet, TransBTS, UNETR, SwimUNETR, Our(WSC-Trans) model actual 3D segmentation effect respectively.

In order to verify the accuracy of our proposed algorithm we choose 3D-Unet[13], 3D-ResUnet[14], Unet++[18], VoxResUnet[46], Vnet[23], 3D-BioUnet[47],VT-Unet[48] , TransBTS[39], UNETR[49], SwimUNETR[50] which have excellent performance in medical image segmentation as deep learning models. In Figure 6, we can clearly observe that Vnet only segmented the cochlea and vestibule, the other three targets were not successfully segmented and the other models successfully completed the corresponding segmentation tasks. By looking at the segmentation results of each model, we can see that the segmentation results of each model are similar in terms of accurate and effective segmentation of the cochlea and vestibule. However, when observing the segmentation of the semicircular canal and facial nerve, it can be seen that the other models have obvious fractures and deformations, where the fracture of the facial nerve mainly occurs in the middle position, and the facial nerve fracture phenomenon is most obvious in the segmentation results of the VoxResUnet model, and in the segmentation results of the semicircular canal, the VT-Unet, TransBTS, UNETR, and SwimUNETR models The segmentation results of the semicircular canal in the bend were smaller or had obvious fractures, and in some models, due to the special structure of the auditory tuberosity, other models than VNet segmented the auditory tuberosity, but the part of the auditory tuberosity connected to the cochlea was too small to be segmented in several models. By visualizing the 3D segmentation results of each model and the real labels, our proposed WSC-Trans model showed better results for segmenting the anatomical structures of the cochlea and facial nerve when compared with the segmentation results of multiple models. The results of the segmentation effects of individual models using multiple evaluation methods are shown in Table II. In the DSS, the facial nerve scores were much lower than the scores of other anatomical structures in the segmentation results of each model, and the difference between the lowest and highest scores of each model was too large 18.69%, and the cochlear segmentation scores were closer to the maximum and minimum differences of 4.32%. In HD95, the difference between the models in cochlear segmentation and vestibular segmentation was very small, except for Vnet, which was also very close, but in the facial nerve segmentation, the difference between the models was very large, with the maximum being 4.5888 and the minimum being 1.1478. In ASSD, similar results were found for cochlear and vestibular segmentation. We believe that this is because the voxels of the cochlea and the facial nerve are different in proportion and the spatial location of the structures is different, and the models focus more on learning simple targets such as the cochlea and the vestibule. By comparing the segmentation results of each model among the four evaluation indexes, it was effectively demonstrated that the segmentation effect of our proposed WSC-Trans model was better than other models in both individual anatomical structure segmentation and overall segmentation, which proved the superiority of the WSC-Trans model.

TABLE II

AVERAGE DSS, HD95, JSS AND ASSD OF THE PROPOSED SEGMENTATION ALGORITHM FOR FIVE TEMPORAL BONE STRUCTURES. STRUCTURES ARE COCHLEA(CO), FACIAL NERVE(FN), OSSICLES(OS), VESTIBULE(VE), AND SEMICIRCULAR CANAL(SE)

|  | DSS(%) ↑ | | | | | | JSS(%) ↑ | | | | | |
|---|---|---|---|---|---|---|---|---|---|---|---|---|
|  | CO | FN | OS | SE | VE | Average | CO | FN | OS | SE | VE | Average |
| 3D Unet | 92.17 | 71.65 | 85.87 | 88.59 | 92.52 | 86.16 | 85.48 | 55.99 | 77.28 | 79.65 | 86.1 | 76.9 |
| 3D ResUnet | 90.07 | 65.68 | 82.83 | 86.62 | 88.16 | 82.67 | 81.94 | 49.12 | 71.35 | 76.5 | 78.88 | 71.56 |
| Unet++ | 92.38 | 69.52 | 91.19 | 89.16 | 92.44 | 86.94 | 85.84 | 53.47 | 83.9 | 80.49 | 85.96 | 77.93 |
| VoxResNet | 92.08 | 69.87 | 90.14 | 88.49 | 92.08 | 86.53 | 85.33 | 53.8 | 82.19 | 79.45 | 85.34 | 77.22 |
| Vnet | 90.62 | - | - | - | 77.71 | - | 82.86 | - | - | - | 63.56 | - |
| 3D BioUnet | 92.03 | 67.22 | 90.15 | 88.28 | 92.23 | 85.98 | 85.24 | 50.76 | 82.19 | 79.07 | 85.59 | 76.57 |
| VT-Unet | 88.21 | 58.32 | 58.75 | 80.28 | 87.98 | 80.11 | 78.92 | 41.62 | 75.24 | 67.28 | 78.57 | 68.32 |
| TransBTS | 90.71 | 67 | 88.4 | 86.87 | 91.38 | 84.87 | 83.02 | 50.63 | 79.35 | 76.95 | 84.18 | 74.83 |
| UNETR | 91.03 | 54.68 | 90.04 | 84.28 | 91.55 | 82.32 | 83.57 | 38.25 | 82.37 | 73.09 | 84.48 | 72.35 |
| Swim UNETR | 92.04 | 58.7 | 90.96 | 88.28 | 92.3 | 84.46 | 85.27 | 41.88 | 83.51 | 79.12 | 85.72 | 75.1 |
| **Ours** | **92.53** | **73.37** | **92.23** | **90.28** | **93.05** | **88.29** | **86.1** | **58.09** | **85.71** | **82.36** | **87.02** | **79.85** |
|  | HD95(Voxel)↓ | | | | | | ASSD(Voxel)↓ | | | | | |
|  | CO | FN | OS | SE | VE | Average | CO | FN | OS | SE | VE | Average |
| 3D Unet | 1 | 1.828 | 1.4116 | 1 | 1 | 1.2479 | 0.0802 | 0.4082 | 0.1824 | 0.1248 | 0.0819 | 0.1755 |
| 3D ResUnet | 1 | 2.9219 | 1.194 | 1.11 | 1.007 | 1.7801 | 0.1021 | 0.5671 | 0.1998 | 0.1596 | 0.1227 | 0.2303 |
| Unet++ | 1 | 1.7379 | 1.2534 | 1.0043 | 1.0043 | 1.4367 | 0.0848 | 0.422 | 0.0963 | 0.1321 | 0.0836 | 0.1638 |
| VoxResNet | 1 | 2.1045 | 1.008 | 1 | 1 | 1.2226 | 0.0817 | 0.4837 | 0.1172 | 0.1393 | 0.0901 | 0.1824 |
| Vnet | 1.2223 | - | - | - | 20.4912 | - | 0.4424 | - | - | - | 3.1614 | - |
| 3D BioUnet | 1 | 2.0358 | 1.0258 | 1 | 1 | 1.2123 | 0.0819 | 0.4209 | 0.1082 | 0.124 | 0.0807 | 0.1631 |
| VT-Unet | 1 | 3.1668 | 1.1427 | 1.055 | 1 | 1.4729 | 0.136 | 0.6941 | 0.1986 | 0.224 | 0.1288 | 0.2763 |
| TransBTS | 1.0043 | 4.321 | 1.3291 | 1.0248 | 1.0043 | 1.7367 | 0.1151 | 0.8151 | 0.2828 | 0.1897 | 0.0979 | 0.3001 |
| UNETR | 1 | 4.5888 | 1.194 | 1.11 | 1.0076 | 1.7801 | 0.0921 | 0.8589 | 0.1459 | 0.2152 | 0.0969 | 0.2818 |
| Swim UNETR | 1 | 2.5343 | **0.9773** | 1 | 1 | 1.3023 | 0.0818 | 0.5364 | 0.1043 | 0.1266 | 0.0819 | 0.1862 |
| **Ours** | **1** | **1.1478** | **0.9874** | **1** | **1** | **1.0931** | **0.0766** | **0.3336** | **0.1008** | **0.1094** | **0.0751** | **0.1391** |

We can observe from Figure 7 that the box plots of the segmentation results of the semicircular canal and the auditory tuberosity show that the maximum and minimum values of the segmentation results of all models except the Vnet model do not differ significantly, but the segmentation results of each model in the segmentation of the auditory tuberosity show outliers, indicating that the segmentation effect of each model on the auditory tuberosity is less stable compared to the segmentation of other targets. The WSC-Trans model showed no outliers in the segmentation of the cochlea, facial nerve and vestibule, and the maximum and minimum segmentation values were higher than those of the other models. Compared with the other models, the difference between the maximum and minimum values of DSS for each segmentation target in the WSC-Trans model was smaller, indicating that the model has stable segmentation ability.

# I. DISCUSSIONS

To select the appropriate parameters in the model multiple sets of experiments were carried out.

In order to better obtain higher-order semantic features in the underlying network using attention mechanisms, as window attention (W), shift window attention (S) and channel attention (C) can be modularized, we made various combinations of the three attention mechanisms, as shown in Table 3, and the segmentation results after various combinations were better than the segmentation results of the comparative model

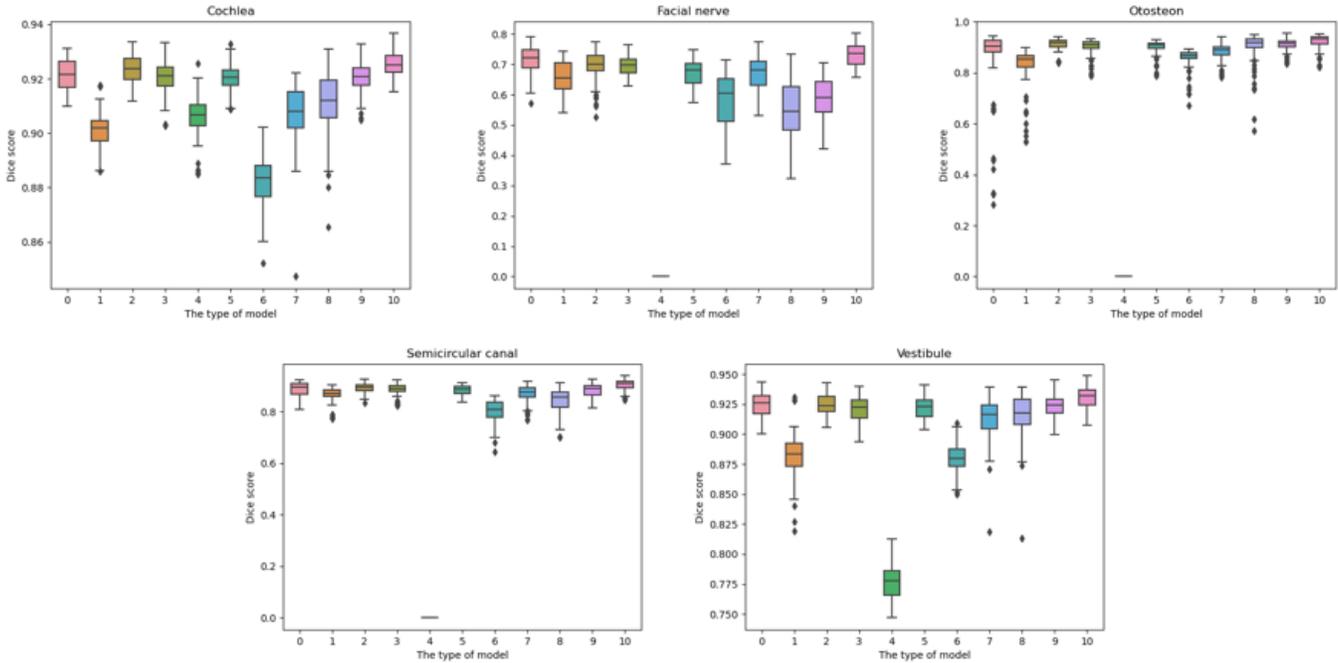

Fig7. Box plots of the DSS results of different models for the cochlea, facial nerve, semicircular canal, auditory tubercle, and vestibular segmentation. Where the numbers 0~10 represent 3D Unet, 3D ResUnet, Unet++, VoxResUnet, Vnet, 3D BioUnet, VT-Unet, TransBTS, UNETR, SwimUNETR, Our (WSC-Trans) model.

proposed in the paper, among which the segmentation results of the WSC combination method were the best.

TABLE III
PERFORMANCE WITH DIFFERENT COMBINATIONS OF ATTENTION MECHANISMS

|  | DSS↑ | JSS↑ | HD95↓ | ASSD↓ |
|---|---|---|---|---|
| WS | 87.8 | 79.11 | 0.1601 | 1.145 |
| WC | 87.89 | 79.36 | 0.1545 | 1.2052 |
| SC | 87.76 | 79.12 | 0.1558 | 1.1574 |
| **WSC** | **88.29** | **79.85** | **0.1391** | **1.0931** |

To give the model better segmentation capability and to take into account the limitations of computational power, we tried to use various Batch sizes and finally a Batch size of 4, as shown in Table 2, gave the model the best segmentation capability.

TABLE IV
PERFORMANCE WITH DIFFERENT BATCH SIZES

|  | DSS↑ | JSS↑ | HD95↓ | ASSD↓ |
|---|---|---|---|---|
| 2 | 87.03 | 78.01 | 0.1644 | 1.1837 |
| **4** | **88.29** | **79.85** | **0.1391** | **1.0931** |
| 6 | 87.74 | 79.01 | 0.1619 | 1.2344 |

References Swim transformer model, we try to use different number of modules to improve the model's effect on advanced semantic features, as shown in Table 5, the model segmentation effect improves as the number of modules increases.

TABLE V
PERFORMANCE WITH DIFFERENT NUMBER OF MODULES

|  | DSS↑ | JSS↑ | HD95↓ | ASSD↓ |
|---|---|---|---|---|
| 1111 | 87.77 | 79.17 | 0.1577 | 1.1701 |
| 1121 | 87.08 | 78.03 | 0.1951 | 1.3046 |
| **1131** | **88.29** | **79.85** | **0.1391** | **1.0931** |

We believe that since as the segmentation target contains many small and complex structures, the proportion of larger target structures increases when the window is enlarged, causing the model to pay more attention to the larger structures and ignore the small and complex ones. Using a smaller window makes the model pay more attention to small structures as well.

TABLE VI
PERFORMANCE WITH DIFFERENT WINDOW SIZES

|  | DSS↑ | JSS↑ | HD95↓ | ASSD↓ |
|---|---|---|---|---|
| 2 | 88.29 | 79.85 | 0.1391 | 1.0931 |
| 3 | 87.89 | 79.26 | 0.1682 | 1.2251 |
| 4 | 87.55 | 78.7 | 0.1728 | 1.2641 |

The cochlea, facial nerve, vestibule, semicircular canal and auditory tuberosity are in different positions in the temporal bone, resulting in different segmentation results. In the evaluation of the segmentation results of the above-mentioned models, we can clearly find that the segmentation effect of the facial nerve is much lower than that of other anatomical tissues. We believe that this is due to the fact that the facial nerve itself is a tubular structure and its own structure is relatively small. The cochlea and vestibular structures are more homogeneous and do not have complex variations, so segmentation is better, while the structures at the junction of the auditory tuberosity and the cochlea are too small to be easily segmented. Therefore, in the design of the model, a newly designed Dense module is used in the coding part to preserve the relevant features as much as possible, and in the next sampling process, the relative position of the features is preserved while expanding the channel, and in the advanced semantic part and the bottom layer of the model, multiple attention mechanisms are used to identify the corresponding target features. In the advanced semantic part and the bottom layer of the model, multiple attention mechanisms are used to identify the corresponding target features, and a small window is considered in window attention and shift window attention to avoid the impact of small target segmentation due to the difference in structure size.

## II. CONCLUSION

In this paper, we propose a semantic segmentation model for temporal bone and nearby biological tissues, which combines the advantages of traditional 3DUnet and Dense modules in the coding design stage to effectively improve the efficiency of image feature extraction, and uses Patch Merging to down sample the image features extracted in the coding part to ensure the effective use of the image features, then uses window In the last layer, we use window attention, shift window attention and channel attention to further extract the target image features. The Res module is used in the decoding part to recover the image features effectively. The framework effectively combines 3D convolution, Transformer and multiple attention mechanisms for the segmentation of relevant biological tissues in temporal bone CT. The final model not only inherits the advantages of using 3D convolution to extract image features and Transformer modeling to connect contextual information, but also utilizes multiple attention mechanisms to learn the global semantic relevance. The segmentation results obtained in this paper on five important anatomical organs are 88.29% average DSC, 79.85% average JSC, 1.0931 average HD95(voxel), and 0.1391 average ASSD (voxel), and our proposed model finally achieves better image

## ACKNOWLEDGMENT

The authors thank the Department of Otolaryngology, Head and Neck Surgery, General Hospital of the Chinese People's Liberation Army for providing technical guidance on the techniques used in this article.